\documentclass[doublecol]{epl2}
\usepackage{amsmath,amssymb}
\usepackage{xcolor}
\newcommand{\lP}{\ell_{\mathrm{P}}}
\newcommand{\SBH}{S_{\mathrm{BH}}}
\newcommand{\CE}{C_E}

\title{One nat per cell}

\author{Ira Wolfson}
\shortauthor{I. Wolfson}

\institute{
  \inst{1} Department of Electronics and Electrical Engineering,
  Braude Academic College of Engineering --- Karmiel, Israel
}

\abstract{
A black hole horizon carries exactly one nat of information per Planck cell,
without naming what the black hole is made of. Two inputs suffice: entropy is
irreducible uncertainty (the configurations no observation can resolve at the
Planck floor, an objective bound, not ignorance), and the horizon is a complete
scrambler, so by the Ehrenfest time its Wigner function has folded below the
floor and no cell can be certified empty. A single cell then holds unresolved
filament crossings; a cell crossed $n$ times admits $n!$ orderings that
scrambling erases, so cardinality $n$ carries weight $1/n!$ and its count is
$\Omega_1=\sum_n 1/n!=e$, one nat. The $N=A/\lP^2$ independent cells give
$\Omega=e^N$, hence $S=N$. Multiplied by the kinematic factor $\tfrac14$ (the
causally accessible fraction of the null phase space, established
separately), this returns $\SBH=A/4\lP^2$. The derivation invokes no field
equations, no microscopic theory, and no posited Hilbert space.
}

\begin{document}
\maketitle

\section{Introduction}

The Bekenstein--Hawking entropy $\SBH = A/4\lP^2$~\cite{Bekenstein73,Hawking75}
poses an explanatory question: what does the horizon count? Every statistical
derivation that reproduces the formula begins by naming a microscopic
theory (string states, spin-network punctures, conformal descendants) and
counting the states that theory supplies; the coefficient follows only after a
theory-specific quantity is tuned to match. The explanation is bought with an
assumption about what the black hole is made of.

This Letter names no microscopic theory. Grant only that the horizon is a
quantum system with a phase space and a causal structure, and that quantum
mechanics holds on it. How much information does one Planck cell of the horizon
carry? The answer is one nat. The area law then follows on multiplying this
per-cell unit by the cell count and by a kinematic factor of $\tfrac14$ that has
an independent origin, established separately from the causal structure of the
null phase space~\cite{WolfsonCQG}. Two commitments carry the argument: a
definition of entropy and one dynamical fact about horizons. Neither names
a constituent.

\section{The horizon at the resolution floor}

Tile the horizon, a topological two-sphere of area $A$, by cells of one Planck
area. Because area is additive, whatever the shapes, the cells sum to $A$ and
there are exactly
\begin{equation}
  N=\frac{A}{\lP^2}
  \label{eq:N}
\end{equation}
of them, with no free normalization constant. The cell is the minimal resolvable phase-space cell, defined by $\Delta q_1\,\Delta p_1\,\Delta q_2\,\Delta p_2\simeq\hbar^2$ and fixed by $[q_i,p_i]=i\hbar$, exact and invariant under unitary (Moyal) evolution. Its configuration-space footprint is the Planck area $\lP^2$; only that area enters the count, not the cell's shape. It does not deform under the scrambling flow: the evolution is dual, the cells deforming with the Wigner function held fixed (the operator picture) or the cells held fixed while $W$ deforms (the state picture), and we use the latter, so the filamentation lives entirely in $W$. There is then no $O(1)$ freedom in $N=A/\lP^2$: a smaller cell would resolve below the floor, a larger one is not the minimal cell, and the equal-area tiling fixes the count by area alone. Quantum mechanics sets no minimal configuration-space area, $[q_1,q_2]=0$, so the $\lP^2$ footprint is the resolution floor, gravitational ($\lP^2=\hbar G/c^3$), not a spatial-commutator effect. The Planck cell is the resolution
floor: no observable distinguishes states that differ only in sub-cell
structure~\cite{Zachos05}. The same boundary between resolvable and sub-floor structure, set by the position--momentum uncertainty, is where Brustein and Kupferman locate the near-horizon entropy divergence~\cite{BrusteinKupferman11}. The horizon cross-section is a two-dimensional
configuration space, so its phase space is four-dimensional; the tiling
\eqref{eq:N} counts the resolvable \emph{configuration} cells, while the
conjugate momenta supply the sub-Planck fine structure that will make a cell's
emptiness uncertifiable. The count below is therefore not a Liouville volume over
$(q,p)$ but a count of resolvable horizon patches, which is why the entropy
scales with area rather than with a phase-space volume.

We take entropy to be irreducible uncertainty: the objective bound on what any
observer can know about the state, fixed by the physical resolution floor and not
by any observer's private ignorance. Formally $S=\log\Omega$, with $\Omega$ the
multiplicity of configurations over the cells no observation can resolve, the
size of the equivalence class of microstates the floor cannot
separate~\cite{Jaynes57,Wehrl78,Safranek19}.

The horizon state is represented by a Wigner quasiprobability $W$ on the horizon
phase space~\cite{Wigner32}, with $\int W\,d\mu=1$ by construction and $W$
allowed to take negative values~\cite{Hillery84}. A cell is \emph{certified
empty} only if a direct measurement at Planck resolution rules it out as
occupied; let $\CE$ be the number of such cells.

\section{Scrambling and the count}

Black holes are the fastest scramblers in nature~\cite{SekinoSusskind08,MSS16}.
Under the chaotic flow the Hamiltonian stretches $W$ into exponentially long
filaments while contracting their transverse thickness~\cite{Zurek01}. The chaotic flow is area-preserving, so a region stretched exponentially along one direction is squeezed exponentially in the transverse one, and repeated folding lays the resulting sub-floor filament back densely across the accessible phase space. Stretching
alone would not suffice (a filament can grow long and still miss regions), but
complete scrambling supplies stretch-and-fold mixing over the full available
phase space, so that by the Ehrenfest time $t_E=\lambda^{-1}\ln(a/\hbar)$ the
folded filament has re-entered the accessible region densely enough that every
Planck patch is reached by Wigner support whose distinguishing structure lies
below the floor. Certifying any such patch empty would require resolving that
sub-cell structure, which the floor forbids. Hence, by $t_E$,
\begin{equation}
  \CE=0:
\end{equation}
no cell is certifiably empty~\cite{HaydenPreskill07,ZurekPaz94}. The horizon
state is now a tangle of unresolved crossings spread over the $N$ cells.

Consider a single cell. The scrambled filament may thread it $0,1,2,\dots$ times. A crossing here is one connected passage of the filamentary support through the cell, registered at the floor; two passages closer than the floor are read as one, not because they merge but because no reading at the floor can separate them, so $n$ is itself a floor-level datum and resolves nothing within the cell.
Walking the filament, these crossings come in a definite order (a finer datum
than their number), but scrambling pushes that order, and the sub-cell positions,
below the floor. The $n!$ orderings of $n$ crossings therefore collapse to one
floor-configuration, so cardinality $n$ carries weight $1/n!$; this Gibbs factor
is not a new postulate but the erased arc-length order, a distinction no
Planck-resolution measurement can recover. These crossings are not identical quanta but distinguishable segments of one connected strand, labelled in principle by their arc-length order and made indistinguishable only by scrambling. This is Maxwell--Boltzmann, and $1/n!$ is exact, not an approximation to a Bose or Fermi count. One symmetry is erased, the arc-length order, so one factor of $1/n!$; a second would quotient a symmetry that is not there. The value $\Omega_1=e$ follows from the count, it is not a target set for it. Nothing weights one crossing number
against another: $n$ is a bare refinement, not a charge, and carries no fugacity.
The cell's compatible class is thus every finite crossing cardinality with its
erased-order weight,
\begin{equation}
  \Omega_1=\sum_{n=0}^{\infty}\frac{1}{n!}=e,
  \label{eq:omega1}
\end{equation}
one nat in a single cell. The constant $e$ is combinatorial (the
exponential-formula count of finite unlabelled crossing collections), not a
thermodynamic input and not an artifact of logarithm base.

After complete scrambling the $N$ cells are counted independently. Scrambling
does not destroy the exact unitary correlations among cells; it hides them below
the floor, where no allowed observation can certify that one cell's crossing
count is tied to another's. Since the entropy counts only floor-certifiable
distinctions, the hidden correlations cannot constrain the coarse-grained count,
and by maximum entropy the joint count factorizes into the independent per-cell
counts~\cite{Jaynes57}. Independence is forced by non-certifiability, not by
dynamical erasure; the ontic correlations remain real and inaccessible, exactly
as the definition of entropy requires. Hence
\begin{equation}
  \Omega=\Omega_1^{\,N}=e^{N},\qquad
  S=\log\Omega=N.
  \label{eq:result}
\end{equation}
One nat per cell.

\section{The area law}

Three ingredients produce the result, none new in itself: the stretch-and-fold
spreading of a Wigner distribution under a mixing Hamiltonian, carried to the
point where no cell is certifiably empty; Boltzmann's reading of entropy as the
logarithm of a multiplicity; and the exponential-formula count $\sum_n 1/n!=e$
per cell. What is new is that these three, assembled at the resolution floor,
already fix the per-cell unit. The area law follows at once. Combined with the
kinematic factor $\tfrac14$, the causally accessible fraction of the null phase
space~\cite{WolfsonCQG},
\begin{equation}
  \SBH=\tfrac14\times N=\frac{A}{4\lP^2}.
\end{equation}
The two factors share no machinery: the counting knows nothing of causal access,
and the causal reduction knows nothing of the counting. The horizon entropy is,
on this account, the combinatorial count of the unresolved cell-crossing
configurations the horizon's own scrambling produces, read at the only resolution
any observer can use.

The construction is theory-free in its counting and its resolution limit; a metric theory of gravity enters at one point only, the conversion of the phase-space cell into a horizon area, set by the curvature coupling of the gravitational action rather than by the field equations. In general relativity that coupling is $G$, the cell is $\lP^2$, and $\SBH=A/4\lP^2$. In an $f(R)$ theory the local coupling is $f'(R)/16\pi G$, the effective cell area is $\lP^2/f'(R)$, and the identical count returns $S=f'(R)A/4\lP^2$, the Wald entropy~\cite{Wald93,IyerWald94,Faraoni10}. The theory-dependence resides entirely in this coupling; the per-cell nat and the kinematic $\tfrac14$ are untouched.

\section{Discussion}

The reading of horizon entropy as inaccessible information is
older~\cite{FrolovNovikov93,EngelhardtWall18}, and the observational-entropy
program makes the coarse-graining precise~\cite{Safranek19}. Sorkin's proposal
that this entropy is entanglement entropy on the horizon at about one bit per
Planckian plaquette~\cite{Sorkin83} is the closest heuristic ancestor of the
per-cell unit derived here, though it posits the bit rather than counting it.
What none of these supply, and what is new, is the coefficient obtained by
counting the scrambled distribution rather than by tracing field modes or
defining a holographic functional. Padmanabhan's horizon equipartition reaches
the same area scaling from $N=A/\lP^2$ surface degrees of
freedom~\cite{PadmanabhanEquip10}, but fixes their content by an equipartition
law rather than by counting unresolved configurations; phase-space
derivations~\cite{Xiao19} reach the exact coefficient by fixing a degeneracy by
hand. A graph-theoretic tiling that enumerates topologically distinct Planck-area configurations likewise recovers the area law and its logarithmic term~\cite{Davidson19}; there the counted objects are the tiling topologies, here the unresolved crossings of the scrambled distribution. Here nothing is tuned, no constituents are named, and no parameter is free:
the count is of the horizon's own scrambled state, read at the Planck floor.

\end{document}